\documentclass[aps,twocolumn,showkeys,preprintnumbers,prl]{revtex4}

\usepackage{graphicx}
\usepackage{dcolumn}
\usepackage{bm}

\begin{document}


\title{Ultrafast electron dynamics in metals}
\author{J.~M.~Pitarke}
\email[Prof.~J.~M.~Pitarke\\
Materia Kondentsatuaren Fisika Saila and Centro Mixto CSIC-UPV/EHU\\
Euskal Herriko Unibertsitatea, 644 Posta Kutxatila\\
E-48080 Bilbo, Basque Country (Spain)\\
Fax: (+34) 94 464 8500\\
E-mail: wmppitoj@lg.ehu.es\\]{}
\author{V.~P.~Zhukov}
\email[Dr.~V.~P.~Zhukov\\
Donostia International Physics Center (DIPC)\\
Manuel de Lardizabal Pasealekua\\
E-20018 Donostia, Basque Country (Spain)\\
Fax: (+34) 94 301 5600\\
E-mail: waxvlvlj@sq.ehu.es\\]{}
\author{R.~Keyling}
\email[Dr.~R.~Keyling\\
Donostia International Physics Center (DIPC)\\
Manuel de Lardizabal Pasealekua\\
E-20018 Donostia, Basque Country (Spain)\\
Fax: (+34) 94 301 5600\\
E-mail: swxkeker@sw.ehu.es\\]{} 
\author{E.~V.~Chulkov}
\email[Prof.~E.~V.~Chulkov\\
Donostia International Physics Center (DIPC) and Centro Mixto CSIC-UPV/EHU\\
Manuel de Lardizabal Pasealekua\\
E-20018 Donostia, Basque Country (Spain)\\
Fax: (+34) 94 301 5600\\
E-mail: waptctce@sq.ehu.es\\]{}
 \author{P.~M.~Echenique}
\email[Prof. P.~M.~Echenique\\
Donostia International Physics Center (DIPC) and Centro Mixto CSIC-UPCV/EHU\\
Manuel de Lardizabal Pasealekua\\
E-20018 Donostia, Basque Country (Spain)\\
Fax: (+34) 94 301 5600\\
E-mail: wapetlap@sq.ehu.es\\]{} 

\date\today

\begin{abstract}
During the last decade, significant progress has been achieved in the rapidly
growing field of the dynamics of {\it hot} carriers in metals. Here we present
an overview of the recent achievements in the theoretical understanding of
electron dynamics in metals, and focus on the theoretical description of the
inelastic lifetime of excited hot electrons. We outline theoretical
formulations of the hot-electron lifetime that is originated in the inelastic
scattering of the excited {\it quasiparticle} with occupied states below the
Fermi level of the solid. {\it First-principles} many-body calculations are
reviewed. Related work and future directions are also addressed.       
\end{abstract}


\maketitle

\section{Introduction}

Recent advances in femtosecond laser technology have made possible the
investigation of electron transfer processes at solid surfaces, which are
known to be the basis for many fundamental steps in surface photochemistry and
ultrafast chemical reactions.\cite{haight,wolf,nienhaus} These are typically
atomic and molecular adsorption processes and catalytic reactions between
different chemical species, which transfer energy from the reaction complex
into the nuclear and electronic degrees of freedom of the solid substrate.
These reactions may induce elementary excitations, such as quantized lattice
vibrations (phonons), collective electronic excitations (plasmons), and
electron-hole (e-h) pairs. In metals, the excitation of e-h pairs leads to an
excited or {\it hot} electron with energy above the Fermi level $\varepsilon_F$
and to an excited or {\it hot} hole with energy below $\varepsilon_F$. It is
precisely the coupling of these hot carriers with the underlying substrate
which governs the cross sections and branching ratios of electronically
induced adsorbate reactions at metal surfaces.  

It is the intent of this Review to discuss the current status of the
rapidly growing field of the dynamics of {\it hot} carriers in metals. The
energy relaxation of these hot electrons and holes is almost exclusively
attributed to the inelastic scattering with {\it cold} electrons below the
Fermi level (e-e scattering) and with phonons (e-ph scattering), since
radiative recombination of e-h pairs may be neglected. Assuming that the
excess energy of the hot carrier is much larger than the thermal energy
$k_BT$, the e-e scattering rate does not depend on temperature.
Furthermore, for excitation energies larger than $\sim 1\,{\rm eV}$
inelastic lifetimes are dominated by e-e scattering, e-ph interactions being
in general of minor importance. Only at energies closer to the
Fermi level, where the e-e inelastic lifetime increases rapidly, does e-ph
scattering become important.\cite{phonons,asier}

Different techniques have recently become available for measuring hot-carrier
lifetimes. Inverse photoemission (IPE)\cite{ipe} and high resolution angle
resolved photoemission (ARPE)\cite{matzdorf} provide an indirect access
to the lifetime of hot electrons and holes, respectively, by measuring the
energetic broadening of transition lines after impigning an electron (IPE) or
a photon (ARPE) into the solid. An alternative to IPE is two-photon
photoemission (2PPE),\cite{fauster} in which a first (pump) photon excites an
electron from below the Fermi level to an intermediate state in the energy
region $\varepsilon_F<\varepsilon<\varepsilon_{\rm vac}$ from where a second
(probe) photon brings the electron to the final state above the vacuum level
$\varepsilon_{\rm vac}$. This technique can also be used to acces the lifetime
of the intermediate state directly in the time domain (time-resolved
2PPE),\cite{petek} by measuring the decrease of the signal as the probe pulse
is delayed with respect to the pump pulse. Recently, it has been demonstrated
that lifetime measurements of electrons and holes in metals can be done by
exploiting the capabilities of scanning tunneling microscopy (STM) and
spectroscopy (STS),\cite{stm1,stm2,stm3} and ballistic electron emission
spectroscopy (BEES) has also shown to be capable of determining hot-electron
relaxation times in solid materials.\cite{bees,claudia} Applications of these
techniques include not only measurements of the scattering rates of hot
carriers in solids, but also measurements of the lifetime of Shockley and
image-potential states at metal surfaces.\cite{review2}

The early theoretical investigations of inelastic lifetimes and mean free
paths of both low-energy electrons in bulk materials and image-potential states
at metal surfaces have been described in previous reviews,\cite{rev1,rev2}
showing that they strongly depend on the details of the electronic band
structure.\cite{rev2} Nevertheless, although accurate measurements of inelastic
lifetimes date back to the mid 1990's, the first band-structure calculations
of the e-e scattering in solids were not carried out until a few years
ago.\cite{campillo1,ekardt,campillo2,campillo3}

Here we outline theoretical formulations of the hot-electron lifetime that is
due to the inelastic scattering of the excited {\it quasiparticle} with
occupied states below the Fermi level. Section 2 is devoted to the study of
electron scattering processes in the framework of time-dependent perturbation
theory. A discussion of the main factors that determine the decay of
excited states is presented in Section 3, together with a review of the
existing theoretical investigations of the lifetime of hot electrons in the
bulk of a variety of metals. The inclusion of exchange-correlation effects and
chemical potential renormalization is addressed in Section 4, and the final
Section presents an overview and future directions.

Unless otherwise is stated, atomic units are used throughout, i.e.,
$e^2=\hbar=m_e=1$. Hence, we use the Bohr radius,
$a_0=\hbar^2/m_e^2=0.529\AA$, as the unit of length and the Hartree,
$H=e^2/a_0=27.2\,{\rm eV}$, as the unit of energy. The atomic unit of velocity
is the Bohr velocity, $v_0=\alpha\,c=2.19\times 10^8\,{\rm cm\,s}^{-1}$,
$\alpha$ and $c$ being the fine structure constant and the velocity of light,
respectively.

\section{Theory}

\begin{figure}
\includegraphics[width=0.95\linewidth]{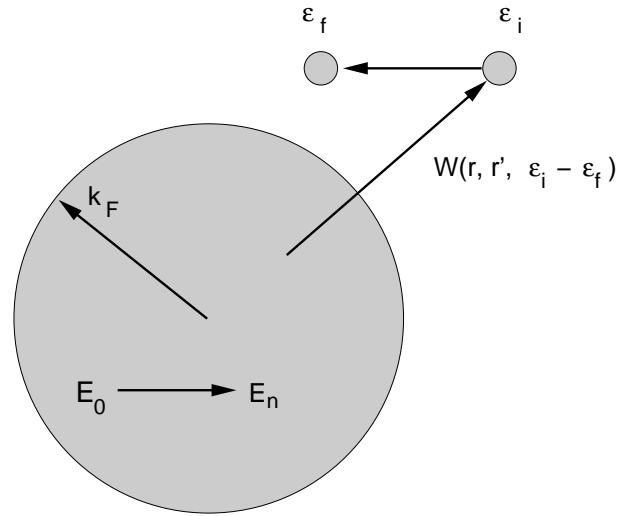}
\caption{\label{fig1}Scattering of an {\it external} excited electron with a
Fermi system of $N$ interacting electrons at $T=0$. The {\it
external} electron in an initial state of energy $\varepsilon_i>\varepsilon_F$
is scattered into an available state of energy $\varepsilon_f$
($\varepsilon_F<\varepsilon_f<\varepsilon_i$) by exciting the {\it cold} Fermi
system from its many-particle ground state of energy $E_0$ to some
many-particle excited state of energy $E_n$
($E_n-E_0=\varepsilon_i-\varepsilon_f$)} 
\end{figure}

We take a Fermi system of $N$ interacting electrons at zero temperature
($T=0$), and consider an {\it external} excited electron interacting with the
Fermi system. Fig.~\ref{fig1} depicts schematically a single inelastic
scattering process for the excited {\it hot} electron. The hot electron in an
initial state $\phi_i(\bf r)$ of energy $\varepsilon_i>\varepsilon_F$ is
scattered into the state $\phi_f(\bf r)$ of energy $\varepsilon_f$
($\varepsilon_F<\varepsilon_f<\varepsilon_i$) by exciting the {\it cold} Fermi
system from its many-particle ground state of energy $E_0$ to some
many-particle excited state of energy $E_n$
($E_n-E_0=\varepsilon_i-\varepsilon_f$). By using the Fermi {\it golden rule}
of time-dependent perturbation theory and keeping only the lowest-order term
in the Coulomb interaction $v({\bf r},{\bf r}' )$ between the {\it hot}
electron and the Fermi gas, the probability $P_{i\to f}$ per unit time for the
occurrence of this process is found to be\cite{pitarke1}
\begin{eqnarray}\label{probability}
P_{i\to f}&=&-2\int d{\bf r}\int d{\bf
r}'\phi_i^*({\bf r})\,\phi_f^*({\bf r}')\,\nonumber\\ &\times&{\rm Im}W({\bf
r},{\bf r}',\varepsilon_i-\varepsilon_f)\phi_i({\bf r}')\,\phi_f({\bf r}),
\end{eqnarray}
where $W({\bf r},{\bf r}',\omega)$ is the so-called screened interaction
\begin{eqnarray}\label{w}
W({\bf r},{\bf r}';\omega)&=&v({\bf r},{\bf r}')+\int d{\bf r}_1d{\bf
r}_2v({\bf r},{\bf r}_1)\nonumber\\
&\times&\chi({\bf r}_1,{\bf r}_2;\omega)v({\bf r}_2,{\bf r}')
\end{eqnarray}
and $\chi({\bf r},{\bf r}';\omega)$ is the density-response function of the
interacting Fermi system.\cite{fetter}

The total decay rate or reciprocal lifetime of the {\it external} excited
electron in the initial state $\phi_i(\bf r)$ of energy $\varepsilon_i$ is
simply the sum of the probabilities $P_{i\to f}$ over all available final
states $\phi_f({\bf r})$ with energies $\varepsilon_f$,
i.e.,
\begin{equation}\label{total}
\tau_i^{-1}=\sum_f\,P_{i\to f},
\end{equation}
where the final states are subject to the condition
$\varepsilon_F<\varepsilon_f<\varepsilon_i$.

The single-particle wave functions and energies $\phi_{i,f}({\bf r})$ and
$\varepsilon_{i,f}$ entering Eq.~(\ref{probability}) can be chosen to be the
eigenfuncions and eigenvalues of an effective Hartree,\cite{fetter}
Kohn-Sham,\cite{gross} or quasiparticle\cite{gunnarsson,nekovee}
hamiltonian.\cite{note2}

\subsection{Non-interacting Fermi sea}

\begin{figure}
\includegraphics[width=0.95\linewidth]{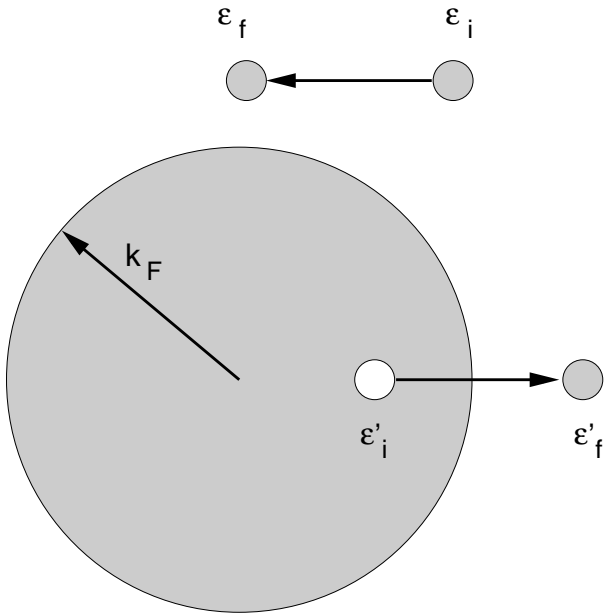}
\caption{\label{fig2}As in Fig.~\ref{fig1}, but now the Fermi system is assumed
to be a system of non-interacting electrons. Hence, single-electron scattering
events are represented, where the {\it external} excited electron of energy
$\varepsilon_i>\varepsilon_F$ is scattered into an available state of energy
$\varepsilon_f$ ($\varepsilon_F<\varepsilon_f<\varepsilon_i$), as in
Fig.~\ref{fig1}, but now carrying one single electron of the Fermi sea from an
initial state of energy $\varepsilon_{i'}$ (below the Fermi level) to a final
state of energy $\varepsilon_{f'}$ (above the Fermi level).}
\end{figure}

If the Fermi sea is assumed to be a system of non-interacting electrons moving
in an effective potential, instead of Eqs.~(\ref{probability})-(\ref{w}) one
first considers the lowest-order probability $P_{i\to f}^{i'\to f'}$ per unit
time for the excited {\it hot} electron in an initial state $\phi_i({\bf r})$
of enery $\varepsilon_i$ to be scattered into the state $\phi_f({\bf r})$ of
energy $\varepsilon_f$ by exciting one single electron of the Fermi sea from an
initial state $\phi_{i'}({\bf r})$ of energy $\varepsilon_{i'}$ to a final
state $\phi_{f'}({\bf r})$ of energy $\varepsilon_{f'}$ (see Fig.~\ref{fig2}),
and then obtains the probability $P_{i\to f}$ by summing over all electrons in
the Fermi sea (below the Fermi level) and all available final
states for these electrons (above the Fermi level):
\begin{equation}\label{non1}
P_{i\to f}=4\pi
\sum_{i',f'}n_{i'}(1-n_{f'})\left|v_{i,i'}^{f,f'}\right|^2
\delta(\varepsilon_i-\varepsilon_f-\varepsilon_{f'}+\varepsilon_{i'}),
\end{equation}
where
\begin{equation}\label{matrix}
v_{i,i'}^{f,f'}=\int d{\bf r}\int d{\bf r}'\phi_i^*({\bf r})\phi_{i'}^*({\bf
r}')\,v({\bf r},{\bf r}')\phi_f({\bf r})\phi_{f'}({\bf r}')
\end{equation}
and $n_i$ are Fermi-Dirac occupation factors, which at $T=0$ are
\begin{equation}
n_i=\theta(\varepsilon_F-\varepsilon_i),
\end{equation}
$\theta(x)$ being the Heaviside step function. The total decay rate
$\tau_i^{-1}$ is obtained by summing over all available final states of the
excited hot electron, as in Eq.~(\ref{total}).

Alternatively, on can simply replace the imaginary part of the interacting
density-response function $\chi({\bf r},{\bf r}';\omega)$ entering
Eq.~(\ref{w}) by its non-interacting counterpart:
\begin{eqnarray}\label{zero}
&&{\rm Im}\,\chi^0({\bf r},{\bf
r}';\omega)=-2\pi\sum_{i',f'}n_{i'}(1-n_{f'})\phi_{i'}^*({\bf
r})\phi_{f'}^*({\bf r}')\nonumber\\
&&\,\,\,\,\,\,\,\,\,\,\,\,\,\,\,\,\,\,\,\,\,\,\,\,\times\phi_{f'}({\bf
r})\phi_{i'}({\bf r}') \delta(\omega-\varepsilon_{f'}+\varepsilon_{i'}).
\end{eqnarray}
Introduction of Eq.~(\ref{zero}) into Eqs.~(\ref{probability})-(\ref{w})
yields again the probability of Eq.~(\ref{non1})

Due to the long-range of the bare Coulomb interaction $v({\bf r},{\bf r}')$
entering Eq.~(\ref{matrix}), this approach yields a total decay rate
$\tau_i^{-1}$ that might be severely divergent, thereby resulting in a lifetime
$\tau_i$ that would be equal to zero. However, many-body interactions of the
Fermi sea, which are fully included in the interacting density-response
function $\chi({\bf r},{\bf r}';\omega)$ entering Eq.~(\ref{w}), are known to
be responsible for a dynamical screening of the Coulomb interaction leading to
a finite lifetime of excited {\it hot} electrons.

Many-body interactions of the Fermi sea are often approximately introduced by
simply replacing the long-range bare Coulomb interaction $v({\bf r},{\bf
r}')$ entering Eq.~(\ref{matrix}) by the frequency-dependent screened
interaction of Eq.~(\ref{w}) at the frequency
$\omega=\varepsilon_i-\varepsilon_f$, although this does not yield in general a
result equivalent to that of Eq.~(\ref{probability}). In the long-wavelength
limit, static ($\omega\to 0$) screening of the Fermi sea can be described by
the Thomas-Fermi
approximation,
\begin{equation}\label{tf}
W^{TF}({\bf r},{\bf
r}')=v({\bf r},{\bf r}')\,{\rm e}^{-q_{TF}|{\bf r}-{\bf r}'|}.
\end{equation}
Here, $q_{TF}$ represents the Thomas-Fermi momentum,
$q_{TF}=(4q_F/\pi)^{1/2}$, $q_F$ being the Fermi momentum.

\subsection{Interacting Fermi sea: random-phase approximation (RPA)}

In an interacting Fermi sea, we need to compute the full interacting
density-response function $\chi({\bf r},{\bf r}';\omega)$ entering
Eq.~(\ref{w}). This function is also known to yield within linear-response
theory the electron density induced in a many-electron system by an external
potential $V^{ext}({\bf r},\omega)$:
\begin{equation}\label{rho1}
\rho^{ind}({\bf r},\omega)=\int d{\bf r}'\chi({\bf r},{\bf
r}';\omega)V^{ext}({\bf r}',\omega).
\end{equation}   

For many years, the dynamical screening in an interacting Fermi system has been
succesfully described in a time-dependent Hartree or, equivalently,
random-phase approximation (RPA).\cite{fetter} In this approach, the
electron density induced by an external potential is obtained as the electron
density induced in a non-interacting Fermi system by both the
external potential $V^{ext}({\bf r},\omega)$ and the potential $V^{ind}({\bf
r},\omega)$ induced by the induced electron density itself
\begin{equation}
V^{ind}({\bf r},\omega)=\int d{\bf r'}v({\bf r},{\bf r}')\rho^{ind}({\bf
r}',\omega).
\end{equation}
Hence, in this approximation
\begin{eqnarray}\label{rho2}
&&\rho^{ind}({\bf r},\omega)=\int d{\bf r}'\left[\chi^0({\bf r},{\bf
r}';\omega)+\int
d{\bf r}_1\right.\nonumber\\ &&\left.\times\int d{\bf
r}_2\chi^0({\bf r},{\bf r}_1;\omega)v({\bf r}_1,{\bf r}_2)\chi({\bf r}_2,{\bf
r}';\omega)\right]V^{ext}({\bf r}',\omega),\nonumber\\
\end{eqnarray}
where $\chi^0({\bf r},{\bf r}';\omega)$ is the density-response function of a
system of non-interacting electrons. Comparing Eqs.~(\ref{rho1})
and~(\ref{rho2}), one finds that in the RPA the interacting density-response
function is obtained from the knowledge of the non-interacting
density-response function by solving the integral equation
\begin{eqnarray}\label{rpa}
\chi({\bf r},{\bf r}';\omega)&=&\chi^0({\bf
r},{\bf r}';\omega)+\int d{\bf r}_1\int d{\bf
r}_2\,\chi^0({\bf r},{\bf r}_1;\omega)\nonumber\\
&\times&v({\bf r}_1,{\bf r}_2)\,\chi({\bf r}_2,{\bf r}';\omega). \end{eqnarray}

\section{Results}

\subsection{General considerations}

The decay rate of excited electrons in an interacting Fermi sea is obtained
from Eqs.~(\ref{probability})-(\ref{total}), the main
ingredients being the hot-electron initial and final states [$\phi_i({\bf
r})$ and $\phi_f({\bf r})$] and the imaginary part of the screened interaction
$W({\bf r},{\bf r}';\omega)$. ${\rm Im}W({\bf r},{\bf r}';\omega)$ contains
both a measure of the probability for creating single-particle and collective
excitations in a many-electron system and a measure of the screening of the
interaction between the hot electron and the Fermi sea. In the case of low
energies, where collective modes cannot be produced, the hot-electron lifetime
is mainly determined by a competition between (i) the coupling of the
initial state $\phi_i({\bf r})$ with available states $\phi_f({\bf r})$ above
the Fermi level, (ii) the phase space available for the creation of
electron-hole (e-h) pairs, and (iii) the dynamical screening of the Fermi sea.

\subsubsection{Coupling with available states above the Fermi level}

The coupling of the excited electron $\phi_i({\bf r})$ with available states
$\phi_f({\bf r})$ above the Fermi level strongly depends on whether the excited
quasiparticle is a bulk or a surface state.

A partially occupied band of Shockley surface states typically occurs in the
gap of free-electron-like $s$,$p$ bands.\cite{shoc1,shoc2,shoc3} Another
class of surface states occurs in the vacuum region of metal surfaces with a
band gap near the vacuum level. These are {\it unoccupied} Rydberg-like
image-potential states, which appear as a result of the self-interaction that
an electron near the surface suffers from the polarization charge it induces
at the surface.\cite{echenique, smith}

\begin{figure}
\resizebox{!}{50mm}{\includegraphics{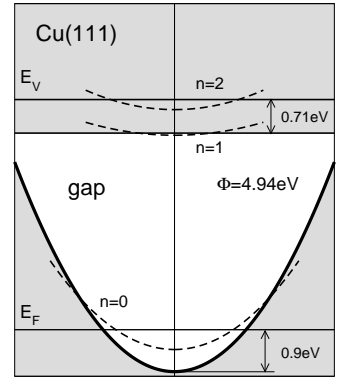}}
\caption{\label{fig3}Schematic representation of the electronic band structure
of the (111) surface of Cu.}
\end{figure}

Fig.~\ref{fig3} shows schematically the projection of the bulk band structure
onto the (111) surface of the noble metal Cu. At the $\bar\Gamma$ point
($k_\parallel=0$), the projected band gap extends from $0.9\,{\rm eV}$ below
to $4.23\,{\rm eV}$ above the Fermi level, in such a way that both a Shockley
($n=0$) and an image ($n=1$) state are supported. 

Shockley surface states are known to be localized near the topmost atomic
layer. However, image states are mainly localized outside the solid.
Therefore, image states are expected to be weakly coupled with available
bulk states above the Fermi level and to live much longer than excited
bulk states with the same energy: while the RPA broadening (or linewidth)
$\tau_i^{-1}$ of a bulk state at the energy of the $n=1$ image state on
Cu(111) ($\varepsilon_i-\varepsilon_F=4.12\,{\rm
eV}$) is found to be $304\,{\rm meV}$, the $n=1$ image-state linewidth is
reduced to $29\,{\rm eV}$.\cite{imanol2}

Coupling of the image state with the crystal occurs through the decay into
bulk unoccupied states lying below the bottom of the projected band gap
(which yields a linewidth of $17\,{\rm meV}$) and also through the decay into
the unoccupied part of the $n=0$ Shockley surface state lying within the
projected band gap (which leads to a linewidth of $12\,{\rm meV}$). A measure
of the coupling of image states to bulk states of the solid can be given
approximately by the penetration of the image-state wave function into the
solid, which in the case of the $n=1$ image state on Cu(111) is found to be of
$22\%$. However, the contribution to the linewidth coming from the decay into
bulk states ($17\,{\rm meV}$) is still well below $0.22$ times the linewidth
of a bulk state at the energy of the image state, i.e., $0.22\times 304\,{\rm
meV}$). This is due to the presence of a projected band gap at the surface,
which in the case of Cu(111) reduces considerably the phase space available
for real transitions of the $n=1$ image state into bulk unoccupied states.
This considerable reduction of the available phase space for real transitions
does not  occur in the case of Cu(100); this explains the known fact that
although the penetration of the $n=1$ image state on Cu(100) is approximately
4 times smaller than in the case of Cu(111) the ratio between the linewidths
of the $n=1$ image state on the (111) and (100) surfaces of Cu is smaller than
2.\cite{imanol2}

\subsubsection{Phase space versus dynamical screening}

In the case of simple metals that can be described by a uniform free-electron
gas of density $n_0$, which is characterized by the density parameter
$r_s=(3/4\pi n_0)^{1/3}/a_0$, a careful analysis of the phase space available
for
the creation of e-h pairs yields a decay rate of hot electrons with energies
near the Fermi level ($\varepsilon_i-\varepsilon_F<<\varepsilon_i$) of the
form\cite{rev2}
\begin{equation}\label{limit1}
\tau_i^{-1}=f(r_s)(\varepsilon_i-\varepsilon_F)^2.
\end{equation}
Furthermore, in the high-density limit ($r_s\to 0$),
Eqs.~(\ref{probability})-(\ref{total}) yield\cite{rev2,qf,note3}
\begin{equation}\label{limit2}
f(r_s)={(3\pi^2/2)^{1/3}\over 36}\,r_s^{5/2}.
\end{equation}
This simple formula shows that the decay rate decreases as the electron
density increases, which is also found to be true in the case of simple and
noble metals ($2<r_s<6$) and hot electrons with energies lying a few
electronvolts above the Fermi level.

As the electron density increases, the density of states (DOS) is known to
increase. Hence, one might be tempted to conclude that as the electron density
increases there are more electrons available for the creation of e-h pairs
near the Fermi level, which would eventually yield an enhanced hot-electron
decay. However, Eq.~(\ref{limit2}) shows that this is not the case. The
reason for this behaviour is twofold. On the one hand, due to momentum and
energy conservation the number of states available for real transitions in
metals is typically weakly dependent on the actual electron density. On the
other hand, as the electron density increases the ability of electrons to
screen the Coulomb interaction with the {\it external} excited electron also
increases, which leads to a smaller screened interaction and a reduced
hot-electron decay.

\begin{figure}
\includegraphics[width=0.95\linewidth]{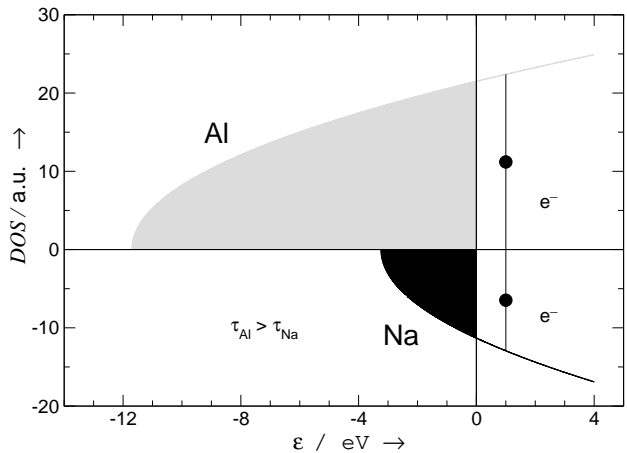}
\caption{\label{fig5}Density of states of a free-electron gas with the
electron density $n_0$ equal to the average density of valence electrons in
Al ($r_s=2.07$) and Na ($r_s=3.99$). The RPA linewidth of high-density Al is
found to be considerably smaller than the corresponding linewidth of
low-density Na.}
\end{figure}

Fig.~\ref{fig5} shows the DOS of a free-electron gas with the electron density
$n_0$ equal to the average density of valence electrons in the simple metals Al
($r_s=2.07$) and Na ($r_s=3.99$), together with the corresponding RPA
linewidths [as obtained from Eqs.~(\ref{probability})-(\ref{total})] of hot
electrons with energy lying $1\,{\rm eV}$ above the Fermi level. While in the
case of Al, with a large DOS near the Fermi level, the linewidth is $14\,{\rm
meV}$, the linewidth of hot electrons with
$\varepsilon_i-\varepsilon_F=1\,{\rm eV}$ in Na is $59\,{\rm meV}$. Hence, hot
electrons live longer in Al than in Na, which is basically due to the strong
screening characteristic of high-density metals like Al.

\subsection{Free-electron gas}

For many years, theoretical predictions of the electron dynamics of bulk
states in solids had been based on a free-electron gas (FEG) or
jellium description of the solid, in which a homogeneous assembly of
interacting electrons is assumed to be immersed in a uniform positive
background. In this model there is translational invariance, the one-particle
states entering Eq.~(\ref{probability}) are momentum eigenfunctions, and
Eqs.~(\ref{probability})-(\ref{total}) are easily found to yield
\begin{equation}\label{uniform1}
\tau_{\bf k}^{-1}=-2\int{d{\bf
q}\over(2\pi)^3}\,{\rm Im}W_{{\bf q},\omega},
\end{equation}
where the energy transfer $\omega=\varepsilon_{\bf k}-\varepsilon_{{\bf k}-{\bf
q}}$ [here, $\varepsilon_{\bf k}=k^2/2$] is subject to
the condition $0<\omega<\varepsilon_{\bf k}-\varepsilon_F$, and
\begin{equation}\label{wnew}
W_{{\bf q},\omega}=v_{\bf q}+v_{\bf q}\,\chi_{{\bf q},\omega}\,v_{\bf q},
\end{equation}
$v_{\bf q}$ and $\chi_{{\bf q},\omega}$ being Fourier transforms of the bare
Coulomb interaction $v({\bf r},{\bf r}')$ and the interacting density-response
function $\chi({\bf r},{\bf r}';\omega)$, respectively. The screened
interaction $W_{{\bf q},\omega}$ is usually
expressed in terms of the inverse dielectric function $\epsilon_{{\bf
q},\omega}^{-1}$, as follows
\begin{equation}
W_{{\bf q},\omega}=v_{\bf q}\,\epsilon_{{\bf q},\omega}^{-1},
\end{equation}
where
\begin{equation}\label{eps}
\epsilon_{{\bf q},\omega}^{-1}=1+\chi_{{\bf q},\omega}\,v_{\bf q}.
\end{equation}

\subsubsection{Non-interacting Fermi sea}

If the Fermi sea is assumed to be a system of non-interacting electrons,
 $\chi_{{\bf q},\omega}$ reduces to the well-known Lindhard
function $\chi_{{\bf q},\omega}^0$,\cite{pines,lindhard}. One can then write
the imaginary part of the screened interaction of Eq.~(\ref{wnew}) as
\begin{eqnarray}\label{one}
{\rm Im}W_{{\bf q},\omega}^0&=&-2\pi\int {d{\bf
k}'\over(2\pi)^3}n_{{\bf k}'}(1-n_{{\bf k}'+{\bf q}})\nonumber\\
&\times&\left|v_{\bf q}\right|^2\delta(\omega-\omega_{{\bf k}'+{\bf
q}}+\omega_{{\bf k}'}).
\end{eqnarray}

\subsubsection{Interacting Fermi sea: RPA}

In the RPA, the interacting density-response function $\chi_{{\bf
q},\omega}^0$ is obtained from Eq.~(\ref{rpa}), i.e.,
\begin{equation}\label{rpanew} \chi_{{\bf q},\omega}=\chi^0_{{\bf
q},\omega}+\chi_{{\bf q},\omega}^0\,v_{\bf q}\,\chi_{{\bf q},\omega},
\end{equation}
or, equivalently [see Eq.~(\ref{eps})],
\begin{equation}
\epsilon_{{\bf q},\omega}=1-\chi^0_{{\bf q},\omega}\,v_{\bf q}.
\end{equation}
Introducing Eq.~(\ref{rpanew}) into Eq.~(\ref{wnew}), one finds
\begin{eqnarray}\label{two}
{\rm Im}W_{{\bf q},\omega}&=&-2\pi\int {d{\bf
k}'\over(2\pi)^3}n_{{\bf k}'}(1-n_{{\bf k}'+{\bf q}})\nonumber\\
&\times&\left|W_{{\bf q},\omega}\right|^2\delta(\omega-\omega_{{\bf k}'+{\bf
q}}+\omega_{{\bf k}'}),
\end{eqnarray}
which is of the form of Eq.~(\ref{one}) with the bare Coulomb interaction
$v_{\bf q}$ replaced by the RPA screened interaction $W_{{\bf q},\omega}$.
Beyond the RPA, ${\rm Im}W_{{\bf q},\omega}$ cannot always be expressed as
in Eq.~(\ref{two}).

We note that in the high-density limit ($r_s\to 0$) and for hot
electrons with energies lying near the Fermi level
($\varepsilon_{\bf k}-\varepsilon_F<<\varepsilon_{\bf k}$), introduction of
Eq.~(\ref{two}) into Eq.~(\ref{uniform1}) yields the decay rate given by
Eqs.~(\ref{limit1}) and~(\ref{limit2}), which will be referred as
$\tau_{QF}^{-1}$.

\begin{figure}
\includegraphics[width=0.95\linewidth]{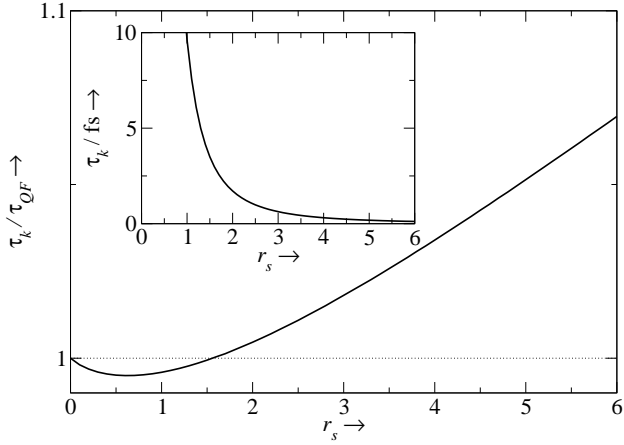}
\caption{\label{fig6}RPA lifetime $\tau_k$ of hot electrons in a free-electron
gas, divided by $\tau_{QF}$ [Eqs.~(\ref{limit1}) and~(\ref{limit2})], as a
function of the electron-density parameter $r_s$ and for hot electrons in the
vicinity of the Fermi surface.}
\end{figure}

\begin{figure}
\includegraphics[width=0.95\linewidth]{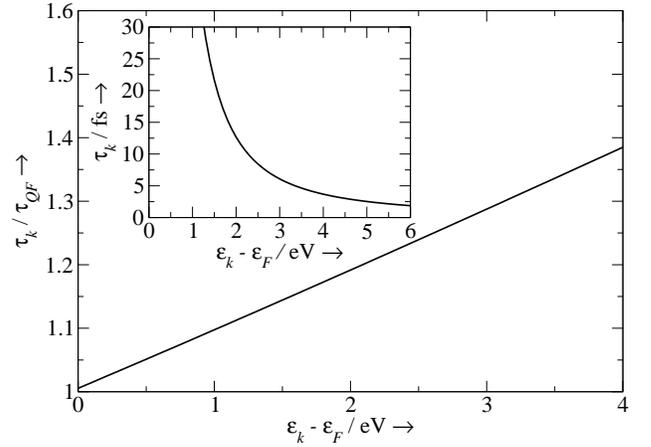}
\caption{\label{fig7}RPA lifetime $\tau_k$ of hot electrons in a
free-electron gas, divided by $\tau_{QF}$ [Eqs.~(\ref{limit1}) and
(\ref{limit2})], as a function of $\varepsilon_k-\varepsilon_F$ for an electron
density equal to that of valence electrons in Al ($r_s=2.07$)}
\end{figure}

Figs.~ 6 and 7 show the RPA lifetime $\tau_{\bf k}$ of hot electrons in a
free-electron gas, divided by $\tau_{QF}$, as a function of the
electron-density parameter $r_s$ for hot electrons in the vicinity of the
Fermi surface (Fig.~\ref{fig6}), and as a function of
$\varepsilon_{\bf k}-\varepsilon_F$ for an electron density equal to that of
valence electrons in Al (Fig.~\ref{fig7}). Although the high-density limit of
Eq.~(\ref{limit2}) only reproduces the full RPA calculation as $r_s\to 0$,
Fig.~\ref{fig6} shows that differences between this high-density approximation
($\tau_{QF}$) and the full RPA calculation ($\tau$) are very small at electron
densities with $r_s<2$ and go up to no more than $7\%$ at $r_s=6$. In the
inset of Fig.~\ref{fig6} the RPA lifetime is represented, showing that as
occurs in the high-density limit [see Eq.~(\ref{limit2})] the RPA lifetime
$\tau_{\bf k}$ increases very rapidly with the electron density.

In the limit $\varepsilon_{\bf k}\to\varepsilon_F$ the
available phase space for real transitions is simply
$\varepsilon_{\bf k}-\varepsilon_F$, which yields the
$(\varepsilon_{\bf k}-\varepsilon_F)^2$ quadratic scaling of
Eq.~(\ref{limit1}). However, as the energy increases momentum and energy
conservation prevents the available phase space from being as large as
$\varepsilon_{\bf k}-\varepsilon_F$. As a result, the actual lifetime
departures from the quadratic scaling predicted for electrons in the vicinity
of the Fermi level (see Fig.~\ref{fig7}), differences between the full RPA
lifetime $\tau_{\bf k}$ and the lifetime $\tau_{QF}$ dictated by
Eqs.~(\ref{limit1}) and~(\ref{limit2}) ranging from $\approx 0.5\%$ at
$\varepsilon_{\bf k}\approx\varepsilon_F$ to $\approx 40\%$ at
$\varepsilon_{\bf k}-\varepsilon_F=4\,{\rm eV}$.

\subsection{Random-$k$ approximation}

For the description of lifetimes and inverse mean free paths (IMFP) in
non-free-electron solids, several authors have employed the so-called
random-$k$ approximation first considered by Berglund and
Spicer\cite{berglund} and by Kane.\cite{kane} The starting point of
this approximation is Eq.~(\ref{non1}) with the bare Coulomb interaction
$v({\bf r},{\bf r}')$ replaced by a frequency-dependent screened interaction
$W({\bf r},{\bf r}';\omega)$. The random-$k$ approximation is then the result
of replacing all the squared matrix elements $\left|W_{i\to f}^{i'\to
f'}(\omega)\right|^2$ by their average $M_i$ over all available states $i'$,
$f'$, and $f$. All the summations entering Eqs.~(\ref{total}) and~(\ref{non1})
can then be replaced by integrations over the corresponding density of states,
and one can write the reciprocal lifetime as
\begin{equation}\label{random}
\tau_i^{-1}={\pi\over
2}\,M_i\,\int_{\varepsilon_F}^{\varepsilon_i}d\varepsilon\,\rho(\varepsilon)
\int_{\varepsilon_F-(\varepsilon_i-\varepsilon)}^{\varepsilon_F}d\varepsilon'\,
\rho(\varepsilon')\rho(\varepsilon_i+\varepsilon'-\varepsilon), \end{equation}
where the quantities $\rho(\varepsilon)$ are assumed to include spin, i.e.,
\begin{equation}
\int_{-\infty}^{+\infty}d\varepsilon\rho(\varepsilon)=N,
\end{equation}
$N$ being the total number of electrons.

A further simplification may be achieved if the density of states is assumed
to take the constant value $\rho_i$ below and above the Fermi level.
Eq.~(\ref{random}) then reduces to
\begin{equation}
\tau_i^{-1}={\pi\over 4}\,\rho_i^3\,M_i^2\,(\varepsilon_i-\varepsilon_F)^2.
\end{equation}
As the electron density increases, both the DOS and the screening of the Fermi
sea increase. Since an enhanced screening yields a reduced $M_i$ factor, the
actual dependence of the hot-electron lifetime on the electron density would
be the result of the competition between DOS and screening effects. Screening
effects typically dominate leading to a lifetime that increases with the
electron density, as occurs in the case of a free-electron gas.

The random-$k$ approximation was first used by Berglund and Spicer in order to
explain experimental photoemission studies of Cu and Ag.\cite{berglund}
Scattering rates of electrons in Si were computed by Kane,\cite{kane} and
Krolikowsky and Spicer\cite{ks} employed the random-$k$ approximation to
calculate the energy dependence of the IMFP of electrons in Cu from the
knowledge of density-of-state-distributions in this material which had been
deduced from photoelectron energy-distribution measurements.

More recently, Penn {\it et al.} used the random-$k$ approximation to analyze
the spin-polarized electron-energy-loss spectra and hot-electron lifetimes in
ferromagnetic Fe, Ni, Co, and Fe-B-Si alloys.\cite{penn1,penn2} The
experimental spin-dependent lifetime of the $n=1$ image state on Fe(110) was
also interpreted by using this approximation.\cite{passek} Drouhin
developed a model to evaluate the scattering cross section and spin-dependent
inelastic mean free path from the knowledge of density-of-state
distributions, and applied it to Cr, Fe, Co, Ni, Gd, Ta, and the noble metals
Cu, Ag, and Au.\cite{drouhin} An analogous approach was developed by Zarate
{\it et al.}, where simple approximations to the DOS were used to obtain
analytical expressions for the electron lifetimes in transition
metals.\cite{zarate}

The random-$k$ approximation has also been discussed and compared to
first-principles calculations by Zhukov {\it et al.}.\cite{zhukov11,zhukov2}
It was shown that when initial and final states are either both $sp$ or both
$d$ states then with a proper choice of the averaged matrix elements $M_i$
Eq.~(\ref{random}) yields lifetimes in reasonable agreement with more
elaborated calculations.

\subsection{First-principles calculations}

First-principles calculations of the scattering rates of hot-electrons in
periodic solids were first carried out only a few years ago by Campillo {\it
et al.}.\cite{campillo1} In this work, hot-electron inverse lifetimes were
obtained from the knowledge of the {\it on-shell} electron self-energy of
many-body theory, which in the so-called $G^0W$ approximation yields exactly
the same result as Eqs.~(\ref{probability})-(\ref{total}) above.

For periodic crystals, the single-particle wave functions entering
Eq.~(\ref{probability}) are Bloch states $\phi_{{\bf k},i}({\bf r})$ and
$\phi_{{\bf k}-{\bf q},f}({\bf r})$ with energies $\varepsilon_{{\bf k},i}$ and
$\varepsilon_{{\bf k}-{\bf q},f}$, $i$ and $f$ representing band indices.
Hence, introducing Eq.~(\ref{probability}) into Eq.~(\ref{total}) and Fourier
transforming one finds
\begin{eqnarray}\label{bands1}
\tau_{{\bf k},i}^{-1}=&-&2\sum_f\int{d{\bf q}\over(2\pi)^3}\sum_{{\bf
G},{\bf G}'}B_{i,f}^*({\bf q}+{\bf G})\nonumber\\ &\times& B_{i,f}({\bf
q}+{\bf G}')\,{\rm Im}W_{{\bf G},{\bf G}'}({\bf q},\omega),
\end{eqnarray}
where the energy transfer $\omega=\varepsilon_{{\bf k},i}-\varepsilon_{{\bf
k}-{\bf q},f}$ is subject to the condition $0<\omega<\varepsilon_{{\bf
k},i}-\varepsilon_F$, the integration is extended over the first Brillouin
zone (BZ), the vectors ${\bf G}$ and ${\bf G}'$ are reciprocal lattice
vectors, $B_{if}$ represent matrix elements of the form
\begin{equation}\label{bands2} B_{i,f}({\bf q}+{\bf G})=\int d{\bf
r}\phi_{{\bf k},i}^*({\bf r})\,{\rm e}^{i({\bf q}+{\bf G})\cdot{\bf
r}}\phi_{{\bf k}-{\bf q},f}({\bf r}),
\end{equation}
and $W_{{\bf G},{\bf G}'}({\bf q},\omega)$ are Fourier coefficients of the
screened interaction $W({\bf r},{\bf r}';\omega)$. As in the case of the
free-electron gas, the Fourier coefficients  $W_{{\bf G},{\bf G}'}({\bf
q},\omega)$ are usually expressed in terms of the inverse dielectric matrix
\begin{equation}\label{bands3} W_{{\bf G},{\bf G}'}({\bf q},\omega)=v({\bf
q}+{\bf G})\,\epsilon_{{\bf G},{\bf G}'}^{-1}({\bf q},\omega),
\end{equation}
$v(\bf q)$ being the Fourier transform of the bare Coulomb interaction $v({\bf
r},{\bf r}')$. In the RPA,
\begin{equation}\label{bands4}
\epsilon_{{\bf G},{\bf G}'}({\bf q},\omega)=\delta_{{\bf G},{\bf
G}'}-\chi_{{\bf G},{\bf G}'}^0({\bf q},\omega)\,v({\bf q}+{\bf G}'),
\end{equation}
where $\chi_{{\bf G},{\bf G}'}^0({\bf q},\omega)$ are Fourier coefficients of
the non-interacting density-response function $\chi^0({\bf r},{\bf
r}';\omega)$.

Couplings of the wave vector ${\bf q}+{\bf G}$ to wave vectors ${\bf q}+{\bf
G}'$ with ${\bf G}\neq{\bf G}'$ appear as a consequence of the existence of
electron-density variations in the solid. If these terms, representing the
so-called crystalline local-field effects, are neglected, one can write
Eq.~(\ref{bands1}) as
\begin{eqnarray}\label{bands5}
&&\tau_{{\bf k},i}^{-1}=2\sum_f{'}\int{d{\bf
q}\over(2\pi)^3}\sum_{{\bf G},{\bf G}'}B_{i,f}^*({\bf q}+{\bf
G})\nonumber\\ &\times& B_{i,f}({\bf q}+{\bf G})\,v({\bf q}+{\bf G}){{\rm
Im}\,\epsilon_{{\bf G},{\bf G}}({\bf q},\omega)\over\left|\epsilon_{{\bf
G},{\bf G}}({\bf q},\omega)\right|^2}.
\end{eqnarray}
This expression accounts explicitly for the three main ingredients entering the
hot-electron decay process. First of all, the coupling of the hot electron
with available states above the Fermi level is dictated by the matrix elements 
$B_{i,f}({\bf q}+{\bf G})$. Secondly, the imaginary part of the dielectric
matrix $\epsilon_{{\bf G},{\bf G}}({\bf q},\omega)$ represents a measure of the
number of states available for the creation of e-h pairs with momentum and
energy ${\bf q}+{\bf G}$ and $\omega$, respectively. Thirdly, the dielectric
matrix in the denominator accounts for the many-body e-e interactions in the
Fermi sea, which dynamically screen the interaction with the external hot
electron.

Since for a given hot-electron energy $\varepsilon$ there are in general
various possible wave vectors and bands, one can define an
energy-dependent reciprocal lifetime by doing an average over wave vectors and
bands. As a result of the symmetry of Bloch states, one finds $\tau^{-1}_{S{\bf
k},i}=\tau^{-1}_{{\bf k},i}$, with $S$ representing a point group symmetry
operation in the periodic crystal. Thus, one can write
\begin{equation}\label{average} \tau^{-1}(\varepsilon)=\frac{\sum_n\sum_{\bf
k}^{\rm IBZ}\,m_{{\bf k},n}\tau^{-1}_{{\bf k},n}}{\sum_n\sum_{\bf k}^{\rm
IBZ}\,m_{{\bf k},n}},
\end{equation}
where $m_{{\bf k},n}$ represents the number of wave vectors ${\bf k}$ lying in
the irreducible element of the Brillouin zone (IBZ) with the same energy
$\varepsilon$.

\begin{figure}
\includegraphics[width=0.95\linewidth]{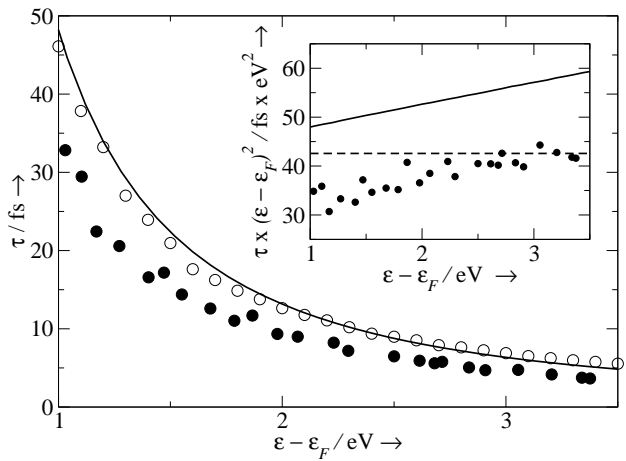}
\caption{\label{fig8} Hot-electron lifetimes in Al, as a function of the
hot-electron energy $\varepsilon$ with respect to the Fermi level
$\varepsilon_F$. Solid circles represent the {\it ab initio} RPA calculation,
as obtained after averaging the lifetime broadening $\tau_{{\bf k},i}^{-1}$ of
Eq.~(\ref{bands1}) over all states with the same energy $\varepsilon$. The
solid line represents the RPA lifetime of hot electrons in a FEG with
$r_s=2.07$. Open circles represent the result obtained from Eq.~(\ref{bands1})
by replacing the hot-electron initial and final states entering the
coefficients $B_{if}$ by plane waves, but with full inclusion of the band
structure in the evaluation of ${\rm Im}W_{{\bf G},{\bf G}'}({\bf
q},\omega)$. The inset exhibits scaled lifetimes of hot electrons in Al. Solid
circles and the solid line represent first-principles and FEG calculations,
respectively, both within RPA. The dashed line represents the prediction of
Eqs.~(\ref{limit1}) and~(\ref{limit2}).}
\end{figure}

\subsubsection{Plane-wave (PW) basis}

The calculations reported in Refs.~\cite{campillo1,campillo2,campillo3} for the
lifetime of hot electrons in the simple metals Al, Mg, and Be, and the noble
metals Cu and Au were carried out from Eqs.~(\ref{bands1})-(\ref{bands4}) by
expanding all one-electron Bloch states in a plane-wave (PW) basis:
\begin{equation}
\phi_{{\bf k},i}({\bf r})={1\over\Omega}\,\sum_{\bf G}u_{{\bf k},i}({\bf
G})\,{\rm e}^{{\rm i}\,({\bf k}+{\bf G})\cdot{\bf r}},
\end{equation}
where $\Omega$ represents the normalization volume. More recently, similar
calculations were reported by Bacelar {\it et al.} for the lifetime of hot
electrons in six transition metals: two fcc metals (Rh and Pd), two bcc metals
(Nb and Mo), and two hcp metals (Y and Ru).\cite{bacelar}

In this approach, one first solves self-consistently for the coefficients
$u_{{\bf k},i}({\bf G})$ the Kohn-Sham equation of density-functional theory
(DFT), with use of the local-density approximation (LDA) for exchange and
correlation\cite{perdew} and non-local norm-conserving ionic
pseudopotentials\cite{pseudo} to describe the electron-ion interaction. From
the knowledge of the eigenfunctions and eigenvalues of the Kohn-Sham
hamiltonian, one can proceed to evaluate the non-interacting
density-response matrix $\chi_{{\bf G},{\bf G}'}^0({\bf
q},\omega)$ (see, e.g., Ref.~\cite{rev2}) and the dielectric matrix
$\epsilon_{{\bf G},{\bf G}'}({\bf q},\omega)$, a matrix equation must then be
solved for the inverse dielectric matrix, and the scattering rate is finally
computed from Eq.~(\ref{bands1}) with full inclusion of crystalline
local-field effects.

Here, we are showing the results of {\it ab initio} RPA calculations of the
average lifetime of hot electrons in the simple metal Al and the noble metals
Cu and Au, all obtained from Eqs.~(\ref{bands1})-(\ref{bands4})
and~(\ref{average}), and we compare these results to the RPA lifetime of hot
electrons in the corresponding FEG as obtained from Eqs.~(\ref{uniform1})
and~(\ref{wnew}). Both {\it ab initio} and FEG calculations were performed
within the very same many-body framework, and the comparison between them 
indicates that while band-structure effects reduce the lifetime of hot
electrons in Al, the presence of non-free-electron-like $d$ bands in the noble
metals considerably enhances the lifetime of hot electrons in Cu and Au.

In Fig.~\ref{fig8}, we show {\it ab initio} (solid circles) and FEG (solid
line) RPA calculations of the average lifetime $\tau(\varepsilon)$ [see
Eq.~(\ref{average})] of hot electrons in the face-centered-cubic (fcc) Al, as
a function of the
hot-electron energy $\varepsilon$ with respect to the Fermi level
$\varepsilon_F$.\cite{campillo1} These calculations show that even
in a free-electron metal like Al band-structure effects play a key role
lowering the hot-electron lifetime by a factor of $\sim 0.65$ for all electron
energies under study. In order to understand the origin of band-structure
effects, an additional calculation is represented by open circles where the
hot-electron initial and final Bloch states entering the coefficients
$B_{i,f}$ of Eq.~(\ref{bands2}) have been replaced by plane waves (as in a
FEG) but keeping the full inverse dielectric matrix of the crystal. This
calculation, which lies nearly on top of the FEG curve, shows that
band-structure effects on both e-h pair creation and the dynamical screening
of the Fermi sea are very small, as occurs in the case of slow
ions.\cite{ions} However, the coupling of hot-electron initial and
final Bloch states appears to be very sensitive to the band structure, which in
the case of Al shows a characteristic splitting over the Fermi level thereby
opening new channels for electron decay and reducing the
lifetime.\cite{campillo2}

Similar results to those exhibited in Fig.~\ref{fig8} for Al were obtained for
the hexagonal closed-packed (hcp) Mg,\cite{campillo2} whose band structure also
splits just above the Fermi level along certain symmetry directions. However,
the splitting of the band structure of this material is not as pronounced as
in the case of Al, and the departure of the hot-electron lifetime in Mg from
the corresponding FEG calculation with $r_s=2.66$ was found to be of about
$25\%$, smaller than in Al.

\begin{figure}
\includegraphics[width=0.95\linewidth]{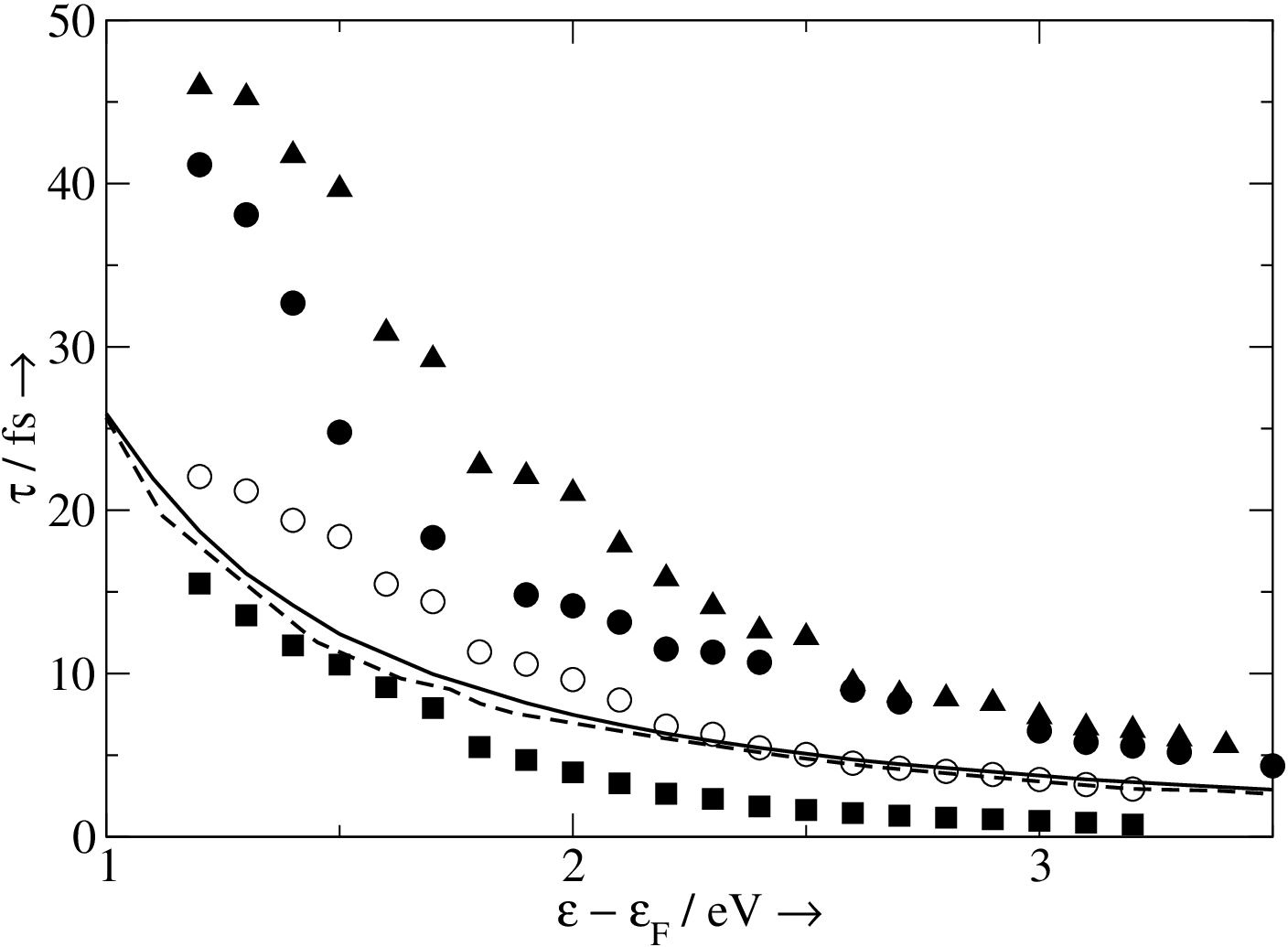}
\caption{\label{fig9}Hot-electron lifetimes in Cu, as a function of the
hot-electron energy $\varepsilon$ with respect to the Fermi level
$\varepsilon_F$. Solid circles represent the {\it ab initio} RPA calculation,
as obtained after averaging the lifetime broadening $\tau_{{\bf k},i}^{-1}$
of Eq.~(\ref{bands1}) over all states with the same energy $\varepsilon$. The
solid line represents the RPA lifetime of hot electrons in a FEG with
$r_s=2.67$. The dotted line represents the {\it ab initio} RPA calculation of
the average lifetime $\tau(\varepsilon)$, but with the $3d$ shell assigned to
the core
in the pseudopotential generation. The squares represent the result of
replacing in Eq.~(\ref{bands5}) all one-electron Bloch states by plane waves
(FEG calculation) but keeping the actual density of states in the evaluation
of ${\rm Im}\epsilon_{{\bf G},{\bf G}'}({\bf q},\omega)$. Open circles
represent the result obtained from Eq.~(\ref{bands5}) by replacing the
hot-electron initial and final states enetring the coefficients $B_{if}$ by
plane waves and the dielectric function in $|\epsilon_{{\bf G},{\bf G}}({\bf
q},\omega)|^{-2}$ by that of a FEG with $r_s=2.67$, but with full inclusion
of the band structure in the calculation of 
${\rm Im}\epsilon_{{\bf G},{\bf G}'}({\bf q},\omega)$. Triangles represent the
result obtained from Eq.~(\ref{bands5}) by replacing the hot-electron initial
and final states entering the coefficients $B_{if}$ by plane waves, but with
full inclusion of the band structure in the evaluation of both
${\rm Im}\epsilon_{{\bf G},{\bf G}'}({\bf q},\omega)$ and 
$|\epsilon_{{\bf G},{\bf G}}({\bf q},\omega)|^{-2}$.}
\end{figure}

\begin{figure}
\includegraphics[width=0.95\linewidth]{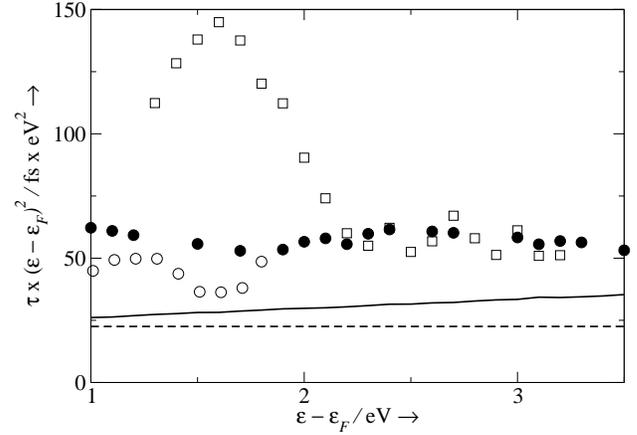}
\caption{\label{fig10}Experimental lifetimes of hot electrons in Cu,
multiplied by $(\varepsilon-\varepsilon_F)^2$ and as a function of
$(\varepsilon-\varepsilon_F)$. The open circles represent the TR-2PPE
measurements reported by Knoesel {\it et al.}\cite{knoesel} for the lifetime of
very-low-energy hot electrons in Cu(111). The TR-2PPE measurements reported
by Ogawa {\it et al.}\cite{ogawa} for the lifetime of hot electrons in Cu(110)
are represented by open squares. For comparison, the {\it ab initio}
calculations
represented in Fig.~\ref{fig9} by solid circles are also represented in this
figure (solid circles), together with the lifetimes of hot electrons in a FEG
with $r_s=2.67$ (solid line) and the corresponding approximation of
Eqs.~(\ref{limit1}) and~(\ref{limit2}) dashed line).}
\end{figure}

\begin{figure}
\includegraphics[width=0.95\linewidth]{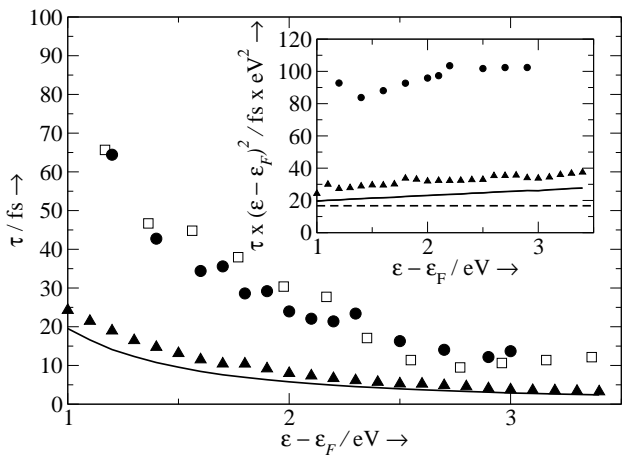}
\caption{\label{fig11}Hot-electron lifetimes in Au, as a function of the
hot-electron energy $\varepsilon$ with respect to the Fermi level
$\varepsilon_F$. Solid circles represent the {\it ab initio} RPA calculation,
as obtained in Ref.~\cite{campillo3} after averaging the lifetime broadening
$\tau_{{\bf k},i}^{-1}$
of Eq.~(\ref{bands1}) over all states with the same energy $\varepsilon$. The
open squares represent the experimental measurements of Ref.~\cite{cao}. The
solid line represents the RPA lifetime of hot electrons in a FEG with
$r_s=3.01$. The triangles represent the result obtained from Eq.~(\ref{bands5})
by replacing the hot-electron initial and final states entering
the coefficients $B_{if}$ by plane waves and the dielectric function in
$|\epsilon_{{\bf G},{\bf G}}({\bf q},\omega)|^{-2}$ by that of a FEG with
$r_s=3.01$, but with full inclusion of the band structure in the calculation
of ${\rm Im}\epsilon_{{\bf G},{\bf G}'}({\bf q},\omega)$. The inset exhibits
scaled lifetimes of hot electrons in Au. The solid circles and the solid line
represent first-principles and FEG calculations, respectively, both within
RPA. The dashed line represents the prediction of Eqs.~(\ref{limit1}) and
(\ref{limit2}). The triangles represent the result obtained from
Eq.~(\ref{bands5}) by replacing the hot-electron initial and final states
entering
the coefficients $B_{if}$ by plane waves and the dielectric function in
$|\epsilon_{{\bf G},{\bf G}}({\bf q},\omega)|^{-2}$ by that of a FEG with
$r_s=3.01$, but with full inclusion of the band structure in the calculation
of ${\rm Im}\epsilon_{{\bf G},{\bf G}'}({\bf q},\omega)$.}
\end{figure}

Figs.~\ref{fig9}-\ref{fig11} exhibit {\it ab initio} (solid circles) and FEG
(solid lines) RPA calculations of the average lifetime $\tau(\varepsilon)$ of
hot electrons in the fcc metals Cu and Au, again as a function
of the hot-electron energy $\varepsilon$ with respect to the Fermi
level. Cu and Au are noble metals with entirely filled $3d$ and $5d$ bands,
respectively. Slightly below the Fermi level, at
$\varepsilon-\varepsilon_F\sim 2\,{\rm eV}$, we have $d$ bands capable of
holding 10 electrons per atom, the one remaining electron being in both Cu and
Au in a free-electron-like band below and above the $d$ bands. Hence, a
combined description of both delocalized $s$ valence bands and localized $d$
bands is needed to address the actual electronic response of these metals. The
results reported in Refs.~\cite{campillo1,campillo2,campillo3} and presented in
Figs.~\ref{fig9}-\ref{fig11}
where found by keeping all $4s^1$ and $3d^{10}$ Bloch states (in the case of
Cu) and all $6s^1$ and $5d^{10}$ Bloch states (in the case of Au) as valence
electrons  in the generation of the pseudopotential.

Also represented in Fig.~\ref{fig9} are additional calculations for the
hot-electron lifetime in Cu, which help to understand the origin and impact
of band-structure effects in this material. These are an {\it ab
initio} RPA calculation similar to the full RPA calculation represented by
solid circles, but with the $3d$ shell assigned to the core (dashed line), and
three approximated calculations from Eq.~(\ref{bands5}) [thus neglecting
crystalline local-field effects] in which the actual band structure is only
considered in (i) the DOS entering the evaluation of ${\rm Im}\,\epsilon_{{\bf
G},{\bf G}}({\bf q},\omega)$ (squares), (ii) the full evaluation of ${\rm
Im}\,\epsilon_{{\bf G},{\bf G}}({\bf q},\omega)$ (open circles), which
also contains the coupling between states below and above the Fermi level
entering the production of e-h pairs, and (iii) the full evaluation of the
imaginary part of the inverse dielectric function ${\rm Im}\,\epsilon_{{\bf
G},{\bf G}}({\bf q},\omega)/\left|\epsilon_{{\bf G},{\bf G}}({\bf
q},\omega)\right|^2$ (triangles).

An inspection of Fig.~\ref{fig9} shows that when the $3d$ shell is assigned to
the core {\it ab initio} calculations (dashed line) nearly coincide with the
FEG prediction (solid line). This must be a consequence of the fact that
band-structure effects in Cu are nearly entirely due to the presence of
localized $d$ electrons. In the presence of $d$ electrons, there are
obviously more states available for the creation of e-h pairs, and one might
be tempted to conclude that $d$ electrons should yield an enhanced
hot-electron decay, especially at the opening of the $d$-band scattering
channel at about $2\,{\rm eV}$ below the Fermi level. Indeed, this is
precisely the result of calculation (i) represented by squares,
which is very close to the calculation reported by Ogawa {\it et
al.}.\cite{ogawa,note4} Nevertheless, a full {\it ab initio} evaluation of
${\rm Im}\,\epsilon_{{\bf G},{\bf G}}({\bf q},\omega)$ yields the calculation
(ii) represented by open circles; this shows that there is no coupling
between $d$ and $sp$ electrons below and above the Fermi level and there is,
therefore, little impact of $d$ electrons on the production of e-h
pairs.\cite{note5} The key role that $d$ electrons play in the hot-electron
decay is mainly due to screening effects. The presence of $d$ electrons gives
rise to additional screening, thus increasing the lifetime of hot electrons
for {\it all} excitation energies, as shown by the result of calculation
(iii) represented by triangles which includes the screening of $d$ electrons
in the {\it ab initio} evaluation of the denominator of Eq.~(\ref{bands5}).
The screening of $d$ electrons is found to increase hot-lectron lifetimes by a
factor of 3. Finally, differences between calculation (iii) (triangles) and
the full {\it ab initio} calculation (solid circles), which are only visible
at hot-electron excitation energies very near the Fermi level, are due to a
combination of band-structure effects on the hot-electron initial and final
$sp$ states above the Fermi level and crystalline local-field effects not
included in Eq.~(\ref{bands5}).

The first time-resolved 2PPE (TR-2PPE) experiments on Cu were performed by
Schmuttenmaer {\it et al.}.\cite{schmu} The electron dynamics on copper
surfaces was later investigated by several
groups.\cite{ogawa,exp1,exp2,exp3,exp4,knoesel} Although there were some
discrepancies among the results obtained in different laboratories, most
measured lifetimes were found to be much longer than those of hot electrons in
a free gas of $4s^1$ valence electrons. The first-principles calculations
reported in Ref.~\cite{campillo1} (see also Fig.~\ref{fig9}) indicate
that this is mainly the result of the screening of $d$ electrons. 

Fig.~\ref{fig10} exhibits (open circles) the TR-2PPE measurements reported by
Knoesel {\it et al.}\cite{knoesel} for the lifetime of very-low-energy
electrons in Cu(111). Since the energy of these electrons is less than $\sim
2{\rm eV}$ above the Fermi level, $d$  electrons (the $d$-band threshold is
located at $\sim 2\,{\rm eV}$ below the Fermi level) cannot participate in the
creation of electron-hole pairs but can participate in the screening of e-e
interactions, thereby increasing the hot-electron lifetime. Fig.~\ref{fig10}
shows that the measurements of Knoesel {\it et al.} are in reasonable
agreement with the first-principles calculations reported in
Ref.~\cite{campillo1} (solid circles).  The relaxation dynamics at the
low-index copper surfaces (111), (100), and (100) were investigated by Ogawa
{\it et al.}\cite{ogawa} in a wider energy range.  Since the experiment
detects the photoemitted electrons in the normal direction (${\bf
k}_\parallel=0$), we have plotted in Fig.~\ref{fig10} (open squares) the
measured lifetimes of hot electrons at the (110) surface of Cu, the only
surface with no band gap for electrons emitted in this direction. At large
electron energies ($\varepsilon-\varepsilon_F>2\,{\rm eV})$, there is very good
agreement between calculated lifetimes (solid circles) and the measurements
reported in Ref.~\cite{ogawa} (open squares). Nevertheless, the calculations
cannot account for the large increase in the measured lifetimes reported in
Ref.~\cite{ogawa} at lower electron energies. The origin of this
discrepancy was discussed by Sch\"one {\it et al.}\cite{sc1}, who
concluded that the increase in the experimentally determined lifetime might be
related to the presence of a photohole below the Fermi level leading to
transient excitonic states.

{\it Ab initio} RPA average lifetimes in Au, as reported in
Ref.~\cite{campillo3}, are represented in Fig.~\ref{fig11}
(solid circles), together with accurate TR-2PPE measurements (open squares)
where the relaxation from electron transport to the surface was expected to be
negligible.\cite{cao} This figure shows that there is very good agreement
between theory and experiment, which are both found to be nearly five times
larger than the lifetime of hot electrons in a free gas of $6s^1$ valence
electrons (solid line). Nevertheless, more accurate linearized augmented
plane-wave (LAPW)\cite{claudia} and plane-wave\cite{idoia} calculations of the
lifetime of hot electrons in Au have been carried out recently, which yield
smaller lifetimes than those reported in Ref.~\cite{campillo3} and represented
in Fig.~\ref{fig11} by an overall factor of $\sim 1.4$. These calculations
have been found to accurately reproduce the BEES spectra for the two
prototypical Au/Si and Pd/Si systems\cite{claudia}.

The triangles of Fig.~\ref{fig11} show the result obtained in
Ref.~\cite{campillo3} from Eq.~(\ref{bands5}) (thus neglecting crystalline
local-field effects) by including the actual band structure of Au in the
evaluation of ${\rm Im}\,\epsilon_{{\bf G},{\bf G}}({\bf q},\omega)$ but
treating the coefficients $B_{i,f}$ and the denominator $\left|\epsilon_{{\bf
G},{\bf G}}({\bf q},\omega)\right|^2$ as in
the case of a FEG with $r_s=3.01$. The result of this calculation is very close
to the FEG calculation (solid lines), which shows
that the opening of the $d$-band scattering channel for e-h pair production
does not play a role. As in the case of Cu, differences between the FEG and
the full {\it ab initio} calculation are mainly a consequence of virtual
interband transitions due to the presence of $d$ electrons, which give rise to
additional screening and largely increase the hot-electron lifetime in Au.

The change $\delta\epsilon$ in the real part of the long-wavelength dielectric
function that is due to the presence of $d$ electrons in the noble metals
is known to be practically constant at the low frequencies
involved in the decay of low-energy hot electrons in these
materials.\cite{eh1,eh2,eh3} Hence, in order to investigate
hot-electron lifetimes and mean free paths in the noble
metals Quinn\cite{quinn63} treated the FEG as if it were embedded in a medium
of dielectric constant $\epsilon_0=1+\delta\epsilon$ instead of unity. The
corrected lifetime is then found to be larger by a factor of
$\epsilon_0^{1/2}$, i.e., $\sim 2.5$ for both Cu and Au. A more accurate
analysis has been carried out recently, in which the actual frequency
dependence of $\epsilon_0$ is included,\cite{aran} showing that for the low
energies of interest the corrected lifetime is still expected to be enhanced
by a factor of $\sim 2.5$ for both Cu and Au. Nevertheless, the {\it ab
initio} calculations exhibited in Figs.~9-11 indicate that the role that
occupied $d$ states play in the screening of $e$-$e$ interactions is much more
important in Au than in Cu.

Finally, we note that the screening of $d$ electrons does not depend on
whether the hot electron can excite $d$ electrons (both in Cu and Au the
$d$-band scattering channel opens at $\sim 2\,{\rm eV}$ below the Fermi level)
or not. Thus, $d$-screening effects do not depend on the hot-electron energy,
and average lifetimes are therefore expected to approximately scale as
$(\varepsilon-\varepsilon_F)^{-2}$, as in the case of a FEG. Nonetheless,
departures
from this scaling behaviour can occur when the hot-electron lifetimes are
measured along certain symmetry directions, as discussed in
Ref.~\cite{campillo1}

\subsubsection{Linear muffin-tin orbital (LMTO) basis}

For noble and transition metals containing $d$ and $f$ electrons it is
sometimes more convenient to perform full all-electron calculations based on
the use of localized basis like linear muffin-tin orbital
(LMTO),\cite{andersen} linear combination of atomic orbital (LCAO),\cite{lcao}
or LAPW basis.\cite{singh} The lifetime of
hot electrons in the noble metals Cu, Ag, and Au, and the $4d$ transition
metals Nb, Mo, Rh, and Pd was determined by Zhukov {\it et
al.}\cite{zhukov1,zhukov2} by using an LMTO basis in the so-called
atomic-sphere approximation (ASA), in which the Wigner-Seitz (WS) cells are
replaced by overlapping spheres.

\begin{figure}
\includegraphics[width=0.95\linewidth]{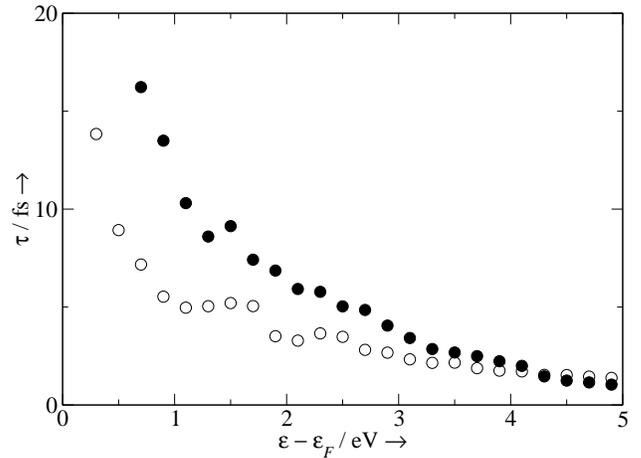}
\caption{\label{fig12}Hot-electron lifetimes in Nb and Pd, as a function of
the hot-electron energy $\varepsilon$ with respect to the Fermi level
$\varepsilon_F$. Solid and open circles represent {\it ab initio} RPA
calculations for Nb and Pd, respectively, as obtained after averaging the
lifetime broadening $\tau_{{\bf k},i}^{-1}$ of Eq.~(\ref{bands1}) over all
states with the same energy $\varepsilon$.}
\end{figure}

In the LMTO method, one-electron Bloch states are expanded as
follows
\begin{equation}
\phi_{{\bf k},i}({\bf r})=\sum_{{\bf R},L}u_{{\bf k},i}({\bf R},L)\,\chi_{{\bf
R},L}({\bf r},{\bf k}),
\end{equation}
where $L=(l,m)$ represents the angular momentum and  $\chi_{{\bf R},L}({\bf
r},{\bf k})$ are the LMTO basis wave functions, which in the ASA are
constructed from the solutions $\varphi_{{\bf R},l}(r)$ to the radial
Schr\"odinger equation inside the overlapping muffin-tin sphere at site ${\bf
R}$ and their energy derivatives. Hot-electron reciprocal lifetimes can
then be evaluated from Eqs.~(\ref{bands1})-(\ref{bands4}) by first solving
self-consistently for the coefficients $u_{{\bf k},i}({\bf R},L)$ the
Kohn-Sham equation of DFT with the use of the LDA for exchange and
correlation\cite{perdew} and then computing the dielectric matrix
$\epsilon_{{\bf G},{\bf G}'}({\bf q},\omega)$ from first principles.

In Fig.~\ref{fig12} we plot the {\it ab initio} RPA calculations that we have
carried out from Eqs.~(\ref{bands1})-(\ref{bands4}) for the average lifetime
$\tau(\varepsilon)$ [see Eq.~(\ref{average})] of hot electrons in the
body-centered-cubic
(bcc) Nb and the fcc Pd.
Nb ($4d^45s^1$) is a transition metal with a partially filled $4d$ band and
one $5s$ valence electron per atom. In this material, the bands of $d$ states
lie in the energy interval from $4\,{\rm eV}$ below to $6\,{\rm eV}$ above the
Fermi level. Hence, contrary to the case of the noble metals where hot
electrons decay by promoting $sp$ electrons from below to above the Fermi
level, in the case of Nb and other transition metals $d$ electrons also
participate in the creation of e-h pairs. Accordingly, the net impact of $d$
electrons in the decay of hot electrons in Nb will be the result of the
competition between the opening of a new scattering channel and the presence
of $d$-electron screening. Valence ($5s^1$) electrons in Nb form a FEG with
electron-density parameter $r_s=3.07$. However, the actual lifetimes of hot
electrons in Nb are considerably shorter than those of hot electrons in the
free gas of $5s^1$ valence electrons, which is obviously due to the presence
of $d$ electrons. Furthermore, if one naively considered a FEG with 5 valence
electrons per atom ($r_s=1.80$), one would predict a lifetime that is $\sim
(3.07/1.80)^{5/2}\sim 4$ times larger [see Eq.~(\ref{limit2})] than in the case
of a FEG with $r_s=3.07$ and nearly 8 times larger than predicted by our {\it
ab initio} calculation. On the one hand, $d$ electrons below and above the
Fermi level in real Nb strongly couple for the creation of e-h pairs, thereby
strongly decreasing the hot-electron lifetime for all electron energies under
consideration. On the other hand, screening of localized $d$ electrons is much
weaker than in the case of free electrons. Hence, while screening effects
dominate in the case of a FEG, the effect of the DOS near the Fermi level
dominates in the case of a gas of electrons with a large number of $d$ states,
which yields the short average lifetimes plotted in Fig.~\ref{fig12}.

Pd ($4d^{10}$) is a transition metal with a filled $4d$ band and no valence
$sp$ electrons. In this material, the bands of $d$ states mainly lie below the
Fermi level, though states above the Fermi level still have a small but
significant $d$ component. The average density of 10 electrons per atom in Pd
is higher than that of 5 electrons per atom in Nb; a FEG picture of the
hot-electron dynamics in these materials would, therefore, lead to longer
lifetimes in Pd. Nevertheless, our {\it ab initio} calculations show that in
the case of low-energy electrons with energies below $4\,{\rm eV}$ this is not
the case, i.e., hot electrons in Pd live shoter than in Nb. This is again due
to the fact that as far as $d$ electrons are concerned the effect of the
density of $d$ states near the Fermi level shortening the lifetime is more
important than the effect of $d$-screening increasing the lifetime, simply
because of the limited ability of $d$ electrons to screen the e-e interactions.
The density of $d$ states near the Fermi level is higher in Pd than in Nb,
thereby leading to lifetimes of low-energy electrons that are shorter in Pd
than in Nb. At energies $\varepsilon-\varepsilon_F>4\,{\rm eV}$ e-h pair
production is restricted to the lowest energies, due to the fact that for the
highest energies $d$ electrons below the Fermi level do not couple with $sp$
electrons at a few eVs above the Fermi level. Thus, at these energies e-h pair
production in Pd is not as efficient as in the case of Nb, which yields
hot-electron lifetimes that are larger in Pd than in Nb.

\section{Green function formalism}

In the theoretical framework of section II we have restricted our analysis to
the case of an {\it external} excited electron interacting with a Fermi system
of $N$ interacting electrons. Nevertheless, a more realistic analysis should
include the excited hot electron as part of one single Fermi system of $N+1$
interacting electrons in which an electron has been added in the one-particle
state $\phi_i({\bf r})$ at time $t'$. In the framework of Green function
theory,\cite{hedin} the probability that an electron will be found in the same
one-particle state at time $t>t'$ is obtained from the knowledge of the
one-particle Green function of a system of $N$ interacting electrons. One
finds,\cite{nekovee}
\begin{equation}
P_i(t',t)=\exp\left[2\,{\rm Im}E_i\,(t-t')\right],
\end{equation}
$\varepsilon_i$ being the so-called quasiparticle energy, i.e., the pole of the
Green function. Hence, the total decay rate or reciprocal lifetime of the
quasiparticle is simply
\begin{equation}\label{life1}
\tau_i^{-1}=-2\,{\rm Im}E_i.
\end{equation}

The quasiparticle energy $E_i$ can be approximately expressed in terms of the
electron self-energy [$\Sigma({\bf r},{\bf r}';E_i)$] and the eigenfunctions
[$\phi_i({\bf r})$] and eigenvalues [$\varepsilon_i$] of an effective
single-particle hamiltonian [$H=-\nabla_{\bf r}^2/2+V({\bf
r})$]:\cite{rev2}
\begin{equation}
E_i\approx\varepsilon_i+\Delta\Sigma(E_i),
\end{equation}
where
\begin{eqnarray}
\Delta\Sigma(E_i)&=&\int d{\bf r}\int d{\bf r}'\phi_i^*({\bf r})
\left[\Sigma({\bf r},{\bf r}';E_i)\right.\cr\cr
&-&\left.V({\bf r})\delta({\bf r}-{\bf r}')\right]\phi_i({\bf r}').
\end{eqnarray}
The total decay rate or reciprocal lifetime of the quasiparticle can,
therefore, be obtained as follows
\begin{eqnarray}\label{tau2}
\tau_i^{-1}&=&-2\,{\rm Im}\Delta\Sigma(E_i)\cr\cr
&=&-2\,\int d{\bf r}\int d{\bf r}'\phi_i^*({\bf r})
{\rm Im}\Sigma({\bf r},{\bf r}';E_i)\phi_i({\bf r}').
\end{eqnarray}

Within many-body perturbation theory,\cite{fetter} it is possible to obtain the
electron self-energy $\Sigma({\bf r},{\bf r}';E_i)$ as a series in the bare
Coulomb interaction $v({\bf r},{\bf r}')$, but due to the long range of this
interaction such a perturbation series contains divergent contributions.
Therefore, the electron self-energy is usually rewritten as a series in the
time-ordered screened interaction $W^{TO}({\bf r},{\bf r}';\omega)$, which can
be built from the knowledge of the Green function itself and coincides for
positive frequencies ($\omega>0$) with the retarded screened interaction of
Eq.~(\ref{w}). In the RPA, the Green function entering the screened
interaction $W^{TO}({\bf r},{\bf r}';\omega)$ is replaced by its
noninteracting counterpart and for positive frequencies one simply finds
Eq.~(\ref{w}) with the density-response function of Eq.~(\ref{rpa}).

To lowest order in the screened interaction, the self-energy $\Sigma({\bf
r},{\bf r}';E_i)$ is obtained from the product of the Green function and the
screened interaction, and is therefore called the $GW$ self-energy. If one
further replaces both the Green function $G({\bf r},{\bf r}';\omega)$ and the
quasiparticle energy $E_i$ entering the GW self-energy by their noninteracting
counterparts [$\Delta\Sigma(E_i)\to\Delta\Sigma^0(\varepsilon_i)$], Eq.
(\ref{tau2}) is easily found to yield an expression for the decay rate which
exactly coincides with the decay rate dictated by
Eqs.~(\ref{probability})-(\ref{total}). This is the so-called {\it on-shell}
$G^0W$ approximation, which within RPA is usually called the {\it on-shell}
$G^0W^0$ approximation. There are various ways of going beyond this
approximation, by either normalizing the electron energy, introducing
short-range xc effects, or by looking at the so-called spectral function
within a non-selfconsistent or a selfconsistent scheme for the Green function.

\subsection{Electron-energy renormalization}

Near the energy shell, i.e., assuming that the deviation of the complex
quasiparticle energy $E_i$ from its noninteracting counterpart $\varepsilon_i$
is small, $E_i$ can be expanded as follows
\begin{equation}
E_i\approx\varepsilon_i+\Delta\Sigma(\varepsilon_i)+(E-\varepsilon_i)\,\left.{\partial\Delta\Sigma(\omega)\over
\omega}
\right|_{\omega=\varepsilon_i}.
\end{equation}
This expansion yields
\begin{equation}\label{zeta0}
E_i\approx\varepsilon_i+Z(\varepsilon_i)\,\Delta\Sigma(\varepsilon_i),
\end{equation}
where $Z$ is the complex quasiparticle weight or renormalization factor:
\begin{equation}\label{zeta}
Z(\varepsilon_i)=\left[1-\left.{\partial\Delta\Sigma(\omega)\over
\omega}\right|_{E=\varepsilon_i}\right]^{-1}.
\end{equation}

\begin{figure}
\includegraphics[width=0.95\linewidth]{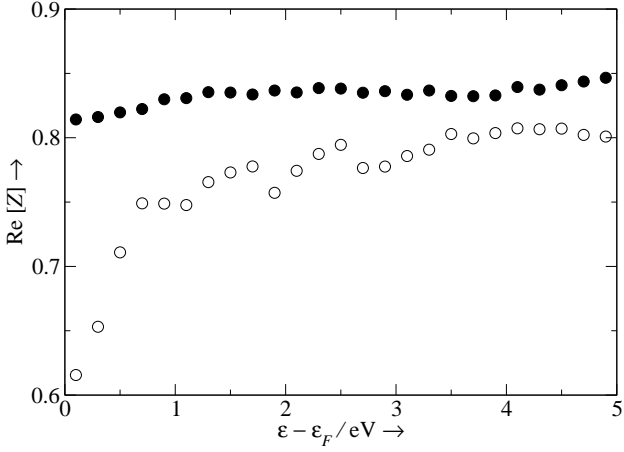}
\caption{\label{fig13}
The real part of the renormalization factor $Z(\varepsilon_i)$ of
Eq.~(\ref{zeta}), versus $\varepsilon_i-\varepsilon_F$, as obtained from a
first-principles evaluation of the $G^0W^0$ electron self-energy in Cu (open
circles) and Pd (crosses).}
\end{figure}

RPA calculations of the renormalization factor $Z(\varepsilon_i)$ of a FEG were
performed by Hedin.\cite{hedin0} These calculations show that in the case of a
free gas of interacting electrons the renormalization factor near the Fermi
level varies from unity in the high-density ($r_s\to 0$) limit to $\sim 0.6$
at $r_s=6$, the imaginary part being very small. We have carried out
first-principles RPA calculations of the renormalization factor of a variety
of real solids. We have found that as in the case of a FEG at metallic
densities ($r_s=2-6$) this factor varies near the Fermi level from $\sim 0.8$
to $\sim 0.6$, as shown in Fig.~\ref{fig13} where the real part of the
renormalization factor $Z(\varepsilon_i)$ of Cu and Pd is represented versus
the noninteracting electron energy $\varepsilon_i$.

Introducing Eq. (\ref{zeta0}) into Eq. (\ref{life1}), one finds
\begin{equation}\label{tau3}
\tau_i^{-1}=-2\left[{\rm Re}Z(\varepsilon_i){\rm Im}\Delta\Sigma(\varepsilon_i)
+{\rm Im}Z(\varepsilon_i){\rm Re}\Delta\Sigma(\varepsilon_i)\right].
\end{equation}
Since the imaginary part of the renormalization factor $Z(\varepsilon_i)$ is
typically very small, Eq.~(\ref{tau3}) yields quasiparticle lifetimes that are
larger than those obtained {\it on-the-energy-shell} by $\sim 30-40\%$.
However, one must be cautious with the use of Eq. (\ref{tau3}) when the
self-energy is calculated in the $G^0W$ approximation by replacing the
electron Green function by its noninteracting counterpart, which is built from
the single-particle wave functions $\phi_i({\bf r})$ and energies
$\varepsilon_i$ In this approximation, it should be more appropriate to
evaluate the self-energy {\it on-shell} by also replacing $E_i$ by
$\varepsilon_i$ in the calculation of the self-energy. This {\it on-shell}
approximation yields a lifetime broadening of the form of Eq. (\ref{tau3})
with $Z(\varepsilon_i)=1$, which exactly coincides with the decay rate
dictated by Eqs.~(\ref{probability})-(\ref{total}).

\subsection{$GW\Gamma$ approximation}

Exchange and short-range correlation of the excited hot electron with the Fermi
system, which are absent in the $G^0W$ approximation, can be included in the
framework of the $GW\Gamma$ approximation.\cite{mahan1,mahan2} In this
approximation, the electron self-energy is of the $GW$ form, but with the
actual time-ordered screened interaction $W^{TO}({\bf r},{\bf r}';\omega)$
replaced by an effective screened interaction
$\tilde W^{TO}({\bf r},{\bf r}';\omega)$ which for $\omega>0$ is obtained as
follows
\begin{eqnarray}\label{w2}
&&\tilde W({\bf r},{\bf r}';\omega)=v({\bf r},{\bf r}')+\int d{\bf r}_1d{\bf
r}_2\left[v({\bf r},{\bf r}_1)\right.\cr\cr
&&+\left.f_{xc}({\bf r},{\bf r}_1;\omega)\right]
\chi({\bf r}_1,{\bf r}_2;\omega)v({\bf r}_2,{\bf r}'),
\end{eqnarray}
the density-response function $\chi({\bf r},{\bf r}';\omega)$ now being
\begin{eqnarray}\label{rpa2}
&&\chi({\bf r},{\bf r}';\omega)=\chi^0({\bf
r},{\bf r}';\omega)+\int d{\bf r}_1\int d{\bf
r}_2\,\chi^0({\bf r},{\bf r}_1;\omega)\cr\cr
&&\times\left[v({\bf r}_1,{\bf r}_2)+f_{xc}({\bf r},{\bf
r}_1;\omega)\right]\,\chi({\bf r}_2,{\bf r}';\omega).
\end{eqnarray}
The kernel $f_{xc}({\bf r},{\bf r}_1;\omega)$ entering Eqs.~(\ref{w2})
and~(\ref{rpa2}), which equals the second functional derivative of the xc
energy functional $E_{xc}[n({\bf r})]$, accounts for the reduction in the
electron-electron interaction due to the existence of short-range xc effects
associated with the excited quasiparticle and with screening electrons,
respectively. In the so-called time-dependent local-density approximation
(TDLDA)\cite{soven} or, equivalently, adiabatic local-density approximation
(ALDA), the {\it exact} xc kernel is replaced by
\begin{equation}
f_{xc}^{ALDA}({\bf r},{\bf
r}';\omega)=\left.{d^2\left[n\varepsilon_{xc}(n)\right]\over
dn^2}\right|_{n=n({\bf r})}\delta({\bf r}-{\bf r}'),
\end{equation}
where $\varepsilon_{xc}(n)$ is the xc energy per particle of a uniform electron
gas of density $n$, and $n({\bf r})$ is the actual electron density at point
${\bf r}$.

Introduction of the $GW\Gamma$ self-energy into Eq. (\ref{tau2}) yields {\it
on-the-energy-shell} ($E_i\to\varepsilon_i$) an expression for the decay rate
of the form of Eqs.~(\ref{probability})-(\ref{total}) with the screened
interaction $W({\bf r},{\bf r}';\omega)$ of Eq.~(\ref{w}) replaced by that of
Eq.~(\ref{w2}). Existing first-principles calculations of hot-electron
lifetimes in solids have all been carried out by neglecting short-range xc
effects, i.e., by taking $f_{xc}({\bf r},{\bf r}';\omega)=0$ in Eqs.~\ref{w2}
and~\ref{rpa2}. First-principles calculations with full inclusion of
short-range xc effects have been performed only very recently.\cite{idoia}

\subsection{Spectral function}

In the framework of many-body theory, the propagation and damping of an excited
electron (quasiparticle) in the one-particle state $\phi_i({\bf r})$ are
dictated by the peaks in the spectral function $A_i(\omega)$, which is closely
related to the imaginary part of the one-particle Green function $G({\bf
r},{\bf r}';\omega)$:\cite{hedin}
\begin{equation}\label{spectral}
A_i(\omega)={1\over\pi}\int d{\bf r}\int d{\bf r}'\phi_i^*({\bf r}){\rm
Im}G({\bf r},{\bf r}';\omega)\phi_i({\bf r}'),
\end{equation}
where
\begin{eqnarray}\label{green}
G^{-1}({\bf r},{\bf r}';\omega)&=&G_0^{-1}({\bf r},{\bf r}';\omega)\cr\cr
&-&\left[\Sigma({\bf r},{\bf r}';\omega)-V({\bf r})\,\delta({\bf r},{\bf
r}')\right].
\end{eqnarray}
The noninteracting Green function $G_0({\bf r},{\bf r}';\omega)$ is obtained
from the eigenfunctions [$\phi_i({\bf r})$] and eigenvalues [$\varepsilon_i$]
of an effective single-particle hamiltonian [$H=-\nabla_{\bf r}^2/2+V({\bf
r})$].

The energetic position of the peak in the spectral function defines the real
part of the quasiparticle energy. The damping rate or lifetime broadening of
the quasiparticle is given by the full width at half maximum (FWHM) $\Delta_i$
of the peak:
\begin{equation}
\tau_i^{-1}=\Delta_i.
\end{equation}
Assuming that the peak in the spectral function $A_i(\omega)$ has a symmetric
Lorentzian form, this definition of the damping rate coincides with the decay
rate dictated by Eq.~(\ref{life1}).

\begin{figure}
\vspace* {1cm}
\includegraphics[width=0.95\linewidth]{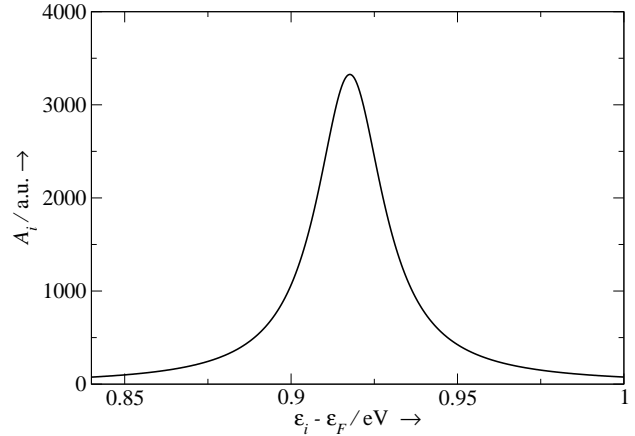}
\caption{\label{fig14}
Spectral function of an excited electron at the $W$ point in Al
($\varepsilon_i=1.02\,{\rm eV}$), versus $\varepsilon_i-\varepsilon_F$, as
obtained from Eq.~(\ref{spectral}) in the $G^0W^0$ approximation.}
\end{figure}

We have carried out first-principles $G^0W^0$ calculations of the spectral
function in Al. For an excited electron at the $W$ point with energy
$\varepsilon_i=1.02{\rm eV}$ above the Fermi level, we have found the spectral
function represented in Fig.~\ref{fig14}. This figure shows that the spectral
function has indeed a single well-defined quasiparticle peak at the
quasiparticle energy $\varepsilon_{qp}=0.92\,{\rm eV}$ above the Fermi level
with the
FWHM $\Delta_i=24\,{\rm meV}$ corresponding to the quasiparticle lifetime
$\tau=27\,{\rm fs}$. Similar calculations were performed by Sch\"one {\it et
al.}\cite{ekardt} for various one-particle Bloch excited states in the simple
metal Al and the noble metals Cu, Ag, and Au. These authors found hot-electron
lifetimes that are in reasonable agreement with the $G^0W^0$ calculations
obtained by Campillo {\it et al.}\cite{campillo1,campillo2,campillo3} from
Eq.~(\ref{tau2}).

Full self-consistent $GW$ self-energy calculations, where the interacting Green
function $G({\bf r},{\bf r}';\omega)$ of Eq.~(\ref{green}) is used
self-consistently to evaluate both the screened interaction and the $GW$
self-energy, have been carried out recently for the FEG\cite{vonbarth} and
simple semiconductors\cite{eguiluz1,eguiluz2}. Although self-consistency is
nowadays known to yield very accurate total energies, existing calculations
tend to indicate that non-self-consistent calculations should be preferred for
the study of quasiparticle dynamics.\cite{vonbarth,eguiluz1} This is due to
the fact that vertex corrections not included in the self-consistent $GW$
self-energy might cancel out the effect of self-consistency, thereby full
self-consistent self-energy calculations that go beyond the $GW$ approximation
yielding results that might be close to $G^0W^0$ calculations.

\section{Summary and Conclusions}

We have presented a survey of current theoretical investigations of the
ultrafast electron dynamics in metals.

First of all, a theoretical description of the finite inelastic lifetime of
excited hot electrons in a many-electron system has been outlined, in the
framework of time-dependent perturbation theory. Then, the main factors that
determine the decay of excited states have been discussed, and the existing
first-principles theoretical investigations of the lifetime of hot electrons
in the bulk of a variety of metals have been reviewed and compared to the
available experimental data. The decay rates obtained within first-order
time-dependent perturbation theory have been shown to coincide with the {\it
on-shell} $G^0W^0$ approximation of many-body theory and to provide a suitable
framework to explain most of the available experimental data for simple and
noble metals.

Finally, various ways of going beyond the $G^0W^0$ approximation have been
discussed, by normalizing the electron energy, introducing short-range xc
effects, or looking at either the $G^0W^0$ or the self-consistent $GW$ spectral
function. First-principles $GW\Gamma$ calculations with full inclusion of
short-range xc effects have been carried out only very recently.\cite{idoia}

Alternatively, nonperturbative treatments of the interaction of {\it external}
excited electrons with a Fermi system have been developed.\cite{nagy,nazarov}
These treatments are based on phase-shift calculations from kinetic
theory\cite{nagy} and a modification of the Schwinger variational principle of
scattering theory,\cite{nazarov} both implemented for electrons in a FEG. The
role of spin fluctuations in the screening and the scattering of excited
electrons in a FEG has also been discussed\cite{nagy2} in the framework of
kinetic theory, with the use of simple physically motivated models.

The experiments considered here all involve very low densities of excited
electrons, and the mutual interaction of excited electrons have been
completely neglected in the theoretical calculations. In a situation in which
the density of excited electrons is high, a different theoretical approach
would be needed. Calculations along these lines have been carried out by
Knorren {\it et al.},\cite{knorren,knorren2} by solving the Boltzmann equation
for carriers in the conduction band.

\begin{acknowledgments}
We acknowledge partial support by the University of the Basque
Country, the Basque Unibertsitate eta Ikerketa Saila, the Spanish
Ministerio de Ciencia y Tecnolog\'\i a, and the Max Planck Research Funds.
\end{acknowledgments}



\begin{thebibliography}{100}
\bibitem{note2} In usual practice all
single-particle wave functions and energies are typically obtained by solving
the single-particle Kohn-Sham equation of density-functional theory in the
so-called local-density approximation (LDA) [see, e.g., Ref.~\cite{gross}].
\end{thebibliography}
\end{document}